\documentclass[showpacs,amssymb,aps,eqsecnum,twocolumn]{revtex4}
\usepackage{amsmath}
\usepackage{graphicx}
\usepackage{color}

\begin{document}
\title{Noise in resistively shunted Josephson junctions}

\author{F. T. Brandt} 
\email{fbrandt@usp.br}
\author{J. Frenkel}
\email{jfrenkel@fma.if.usp.br}
\affiliation{Instituto de F\'{\i}sica, Universidade de S\~ao Paulo,\\
Rua do Mat\~ao, Travessa R, 187, S\~ao Paulo, SP 05508-090, Brazil}
\author{J. C. Taylor}
\email{J.C.Taylor@damtp.cam.ac.uk}
\affiliation{DAMTP, Centre for Mathematical Sciences,
University of Cambridge,\\
Wilberforce Road, Cambridge, CB3 0WA, United Kingdom}

\begin{abstract}
We investigate the dynamics of a resistively shunted Josephson junction. We compute the
Josephson frequency and the
generalized impedances for a variety of the parameters, particularly with relevance to
predicting the measurable effects of zero-temperature current noise in the resistor. 
\end{abstract}

\maketitle

\section{Introduction}\label{sec1}

An experiment reported in 
1981 \cite{koch} measured quantum noise in a resistively shunted
Josephson junction. The high frequency noise (both thermal and quantum) in the
resistor is mixed down to measurable frequencies by the non-linearity of the Josephson
circuit.  This experiment raised some important questions of principle. Quantum noise
is not directly measurable, but it may have measurable effects in non-equilibrium
situations \cite{gavish}. The experiment has, controversially, been related to the question of
``dark energy'' \cite{beck}.  The interpretation of the experiment involves a quantum Langevin
equation (Eq. \eqref{4} below), and the status of this equation has
been critically analyzed by one of us \cite{taylor}. 
{
The existence of such an equation has been demonstrated but only in a simple independent-oscillator model \cite{ford}.

In the experiment \cite{koch}, a resistor $R$ is put in parallel with a Josephson junction
at temperature $T$. The critical current for the junction is $I_0$ and a bias current $I>I_0$
is applied. The voltage across the junction is $V$, and noise fluctuations in this voltage
are measured. The junction has capacitance $C$. Four different junctions were used
with slightly different values of the parameters. As an example, for
junction 2,  $C=0.8$ pF, $I_0=0.5$ mA,
$I=1$ mA, $R=0.6$ $\Omega$, and one temperature for which details are
given was $T=4.2$ K. The noise is measured at frequency 183 kHz.

The theory of the Josephson circuit is particularly simple if the capacitance $C$
can be neglected. In \cite{koch} a simple assumption is made about the dependence on
$C$. One of the motivations for this paper is to test the validity of this assumption.
We find significant  deviations for some values of the voltage.
Our results should be useful if experiments of a similar kind are
performed in the future. }

Our purpose here is to study the solutions of this Langevin equation, and to derive the Josephson frequency and the
generalized impedances {
and hence predict the voltage
noise strength }.
In the interpretation of the experiment \cite{koch}, certain
approximations were made, and we examine the validity of these approximations.
We use two forms of perturbation theory, and also numerical methods.

A mechanical model for the dynamics of the Langevin equation (equation \eqref{4} below) has
been studied experimentally and theoretically in \cite{coullet}.
 
The fundamental theoretical input  \cite{koch2} is  the quantum version of the Nyquist
fluctuation-dissipation theorem (originally derived in \cite{callen}). In the present
context, it relates the current noise in the resistor at temperature $T$ to the conductance $1/R$,
and states
\begin{equation}
\label{1}
\frac{1}{2}\left[S_I(\Omega)+S_I(-\Omega)\right]
=\left(\frac{2\hbar\Omega}{R}\right)\coth\left[\frac{\hbar\Omega}{2T}\right]
\end{equation}
where
\begin{equation}
\label{2}
S_I(\Omega)\delta(\Omega-\Omega')=\langle I_n(\Omega)I_n(-\Omega')\rangle
\end{equation}
$I_n(\Omega)$ being the Fourier transform of the current noise $I_n(t)$ in the resistor, and the expectation value $\langle \dots \rangle$ is with respect to a thermal distribution at temperature $T$
(where we use units such that Boltzmann's constant is 1).

The right hand side of \eqref{1} has the finite limit
\begin{equation}
\label{3}
\frac{2\hbar\Omega}{R}
\end{equation}
as $T\rightarrow 0$, and this offers the possibility to measure quantum noise. For this it is essential
to have the symmetrized version of $S_I$; the unsymmetrized $S_I(\Omega)$ itself has a vanishing low-temperature
limit for $\Omega >0$. This reflects the impossibility of getting energy directly out of
the vacuum. It is, of course, essential that $I_n$ is a quantum, non-commuting, operator.

The right hand side of \eqref{1} contains the distribution function for 
a quantum oscillator of frequency $\Omega$  at
temperature $T$. However, as emphasized in \cite{branchina}, this by no means implies
that the resistor contains oscillators, or even that the system is bosonic.

The second theoretical input is {
the assumed }
quantum Langevin equation, connecting the
voltage and voltage noise to the current and current noise in the shunted Josephson
circuit:
\begin{equation}
\label{4}\left(\frac{\hbar C}{2e}\right)\ddot{\theta}+\left(\frac{\hbar}{2eR}\right)\dot{\theta}+I_0\sin\theta=I+I_n
\end{equation}
where $\theta$ is the Josephson phase (including noise) and $C$ is the capacitance of the junction.
Since $I_n$ is a quantum operator, $\theta$ must be also. The status of this equation is less
secure than that of \eqref{1} (see \cite{taylor}).   The voltage $V$ across the junction is given by
\begin{equation}
\label{5}
V(t)=\frac{\hbar}{2e}\dot{\theta}.
\end{equation}
We will always assume that $I>I_0$. Then,
in the absence of the noise $I_n$, the solutions $\theta_c(t-\hat{t})$ of \eqref{4} settle down, after the decay of transients,
to a periodic dependence on time, with frequency $\Omega_J=2\pi/T_J$, so that
\begin{equation}
\label{6}
\theta_c(t+T_J-\hat{t})=\theta_c(t-\hat{t})+2\pi
\end{equation}
for all $t$. The arbitrary constant $\hat{t}$ is included  in order to make explicit the infinite
set of solution depending on the initial condition (although  \eqref{4} is a second order equation,
the initial value of $\dot{\theta}$ is not an independent initial condition for the steady periodic
solution).

On the right of \eqref{4}, the current noise $I_n$ is certainly a quantum operator, but $I$ is presumably
a classical quantity. Since $I_n$ is small, we can write
\begin{equation}
\label{7}
\theta=\theta_c +\theta_n
\end{equation}
where $\theta_c$ is the (classical) solution and $\theta_n$ is the small quantum noise.
Then the latter obeys, to a good approximation, the linear quantum equation
\begin{equation}
\label{8}
\left(\frac{\hbar C}{2e}\right)\ddot{\theta}_n+\left(\frac{\hbar}{2eR}\right)\dot{\theta}_n+I_0(\cos\theta_c)\theta_n=I_n\end{equation}
in which $\cos\theta_c$ is a $c$-number coefficient.   

Because of \eqref{6}, there is a Fourier series (we use $j$ for $\sqrt{-1}$)
\begin{eqnarray}
\label{9}
\cos\theta_c(t-\hat{t})&=&\sum_k g_k\exp(jk\Omega_J t)\exp(-jk\hat{t}\Omega_J)
\nonumber \\
&\equiv& \sum_k g_k(\hat{t})\exp(jk\Omega_J t)
\end{eqnarray}
with $k$ an integer and $g^*_k=g_{-k}$. 
We define a hybrid Fourier integral/series by
\begin{equation}
\label{10}
I_n(t)=\sum_k \int_0^{\Omega_J} d\Omega I_{nk}(\Omega)\exp[jt(\Omega+k\Omega_J)]
\end{equation}
and similarly for $\theta$ and $V=({\hbar}/{2e})\,{d\theta}/{dt}$.
Then the transform of \eqref{8} is
\begin{eqnarray}
\label{11}
&\sum_{k'}[F_k(\Omega)\delta_{k,k'}+I_0g_{k-k'}(\hat{t})]\theta_{nk'}
\nonumber \\
&\equiv  \sum_{k'}X_{k,k'}(\Omega)\theta_{nk'}(\Omega)
= I_{nk}(\Omega).
\end{eqnarray}
where
\begin{eqnarray}
\label{12}
F_k(\Omega)&=&-\left(\frac{\hbar C}{2e}\right)(\Omega+k\Omega_J)^2
\nonumber \\
&+&j\left(\frac{\hbar}{2eR}\right)(\Omega+k\Omega_J).
\end{eqnarray}
Eq. \eqref{11} has solution
\begin{equation}
\label{13}
\theta_{nk}(\Omega)=\sum_{k'}Y_{kk'}(\Omega) I_{nk}(\Omega)
\end{equation}
where the matrix $Y$ is the inverse of $X$ (for each $\Omega$).

We assume that there is a non-zero limit
\begin{equation}
\label{14}
\lim_{\Omega\rightarrow 0}[j\Omega Y_{0,k}(\Omega)]\equiv \frac{2e}{\hbar} Z_k.
\end{equation}
so that (by \eqref{5})
\begin{equation}
\label{15}
 V_{n0}(0)=\sum_k Z_k I_{nk}(0).
 \end{equation}
This is certainly true for $C=0$ \cite{levinson}, and also consistent with our conclusions in the
remainder of this paper. 

We have not made explicit the $\hat{t}$ dependence of $X$, $Y$ and $Z$, but it follows from
\eqref{9}, \eqref{10}, \eqref{13} and \eqref{14} that  $Z$ is proportional to $\exp(-jk\hat{t}\Omega_J)$.

We define a periodic function $\tilde{I}_n(t)$ associated with $I_n$ by
\begin{equation}
\label{16}
\tilde{I}_n(t)=\sum_k I_{nk}(\Omega=0)\exp(jtk\Omega_J)
\end{equation}
and similarly for $\tilde{V}_n$
Then, for the purpose of computing the $Z_k$ defined in \eqref{14} it is sufficient to use $\tilde{I}_n$ and
$\tilde{V}_n$.

{

The basic problem is to obtain from \eqref{1} and \eqref{8} information about the voltage noise (using \eqref{5}).
One approach to this problem  \cite{koch2,voss} is to simulate the current noise numerically, but it is not
clear how to do this consistently with the frequency dependence in
\eqref{1}. Also, the noise is
supposed to be a quantum operator. 
Our method is indirect. We first obtain the impedances $Z_k$ defined
in \eqref{15}. Since these are assumed to be classical quantities 
(this is the consequence of the assumed existence of a quantum
Langevin equation with {\it classical} coefficients in it), to find them it is sufficient to take the noise in \eqref{8} to
be classical. Having found the impedances $Z_k$, we can use \eqref{15}
(with quantum noise)  together with \eqref{1} to determine the strength 
of the (quantum) voltage noise.
An advantage of this method is that \eqref{1} is exactly respected.

}

Thus, the purpose of this paper is to study the solutions of equation \eqref{4} and \eqref{8}, and the calculation of
$\Omega_J$ and $Z_k$.
This enables us to check the validity of some of the approximations made in \cite{koch}. In section \ref{sec3}, we use perturbation theory for small $C$. This turns out to be a series in $C^2$,
and we are  able to calculate the $O(C^2)$ contribution to $\Omega_J$, and 
 to $Z_k$ but for $k>1$ only. In section \ref{sec5}, we use perturbation theory in $1/I$. Again, there are only even terms,
but we calculate  $O(1/I^2)$ contributions to $\Omega_J$ and to $Z_k$. 
But this approximation is not
useful for the values of $I$ in the experiments. In section \ref{sec6}, 
we give results of numerical
calculations of $\Omega_J$, $R_D$ and $Z_1, Z_2$, for several values of $\beta$ and $i$
(dimensionless parameters defined in \eqref{17} and \eqref{21} below). 
We differ from some other numerical work \cite{koch2,voss}
in that we do not attempt to simulate the actual noise (which should be consistent with \eqref{1}),  but just determine the $Z_k$ factors. Our conclusions are given in graphs, which may be useful in the interpretation of experiments.

One particular conclusion is that, in the range $0<\beta<0.2$ 
(with $\beta$ defined as in \eqref{17} below) $Z_1$ decreases and $R_D$ increases. 
{
In Table \ref{tab2} we show by what factors the values of $Z_1^2$ and of $(Z_1/R_D)^2$
are predicted to change between $\beta=0$ and $\beta=0.38$. }
As we shall discuss,  {
these ratios are new corrections to the predicted voltage noise.}

{
We conclude that the apparent  quantitative agreement between theory and observation in \cite{koch}
may not be as good as it appears. Certainly, our calculations should be relevant to any future
repetition of this type of experiment. For another possible applications of equation \eqref{4}, see Ref. \cite{anna}.

}

\section{Notation and a simple solution}\label{sec2}

It is useful to define the dimensionless variables
\begin{eqnarray}
\label{17}
&i = \dfrac{I}{I_0}, \; v=\dfrac{V}{I_0 R},\;\tau=\dfrac{2eRI_0}{\hbar}t,
\nonumber \\
&\beta=\dfrac{2eR^2CI_0}{\hbar},\; \omega=\dfrac{\hbar}{2eRI_0}\Omega, 
\; z_k=\dfrac{Z_k}{R}.
\end{eqnarray}
In terms of these variables, and denoting $d\theta/d\tau$ by $\theta'$, 
equation \eqref{4} becomes in the absence of the current noise
\begin{equation}
\label{18}
\beta\theta''+\theta'+\sin\theta=i.
\end{equation}

In the special case when $\beta$ is negligible, 
this equation is easily soluble. A particularly
convenient form of a solution is
\begin{eqnarray}
\label{19}
\sin\theta_0&=&\dfrac{1+i\sin(\omega_{J0}\tau)}{i+\sin(\omega_{J0}\tau)},\,\,
\nonumber \\
\cos\theta_0&=&\dfrac{\omega_{J0}\cos(\omega_{J0}\tau)}{i+\sin(\omega_{J0}\tau)},\,\,
\nonumber \\
v_0=\dfrac{d\theta_0}{d\tau}&=&\dfrac{\omega_{J0}^2}{i+\sin(\omega_{J0}\tau)},
\end{eqnarray}
where in general $\omega_J$ is the natural frequency of the solutions of \eqref{18} (when transient
decaying terms have died out), and $\omega_{J0}$ is the value in this approximation
($\beta\simeq 0$):
\begin{equation}
\label{20}
\omega_{J0}=\sqrt{i^2-1}.
\end{equation}
Similarly, $\theta_0$ denotes the solution of \eqref{18} when $\beta=0$.
In this case, equation \eqref{14} gives only three non-zero  values of $Z_k$, with $k=0,\pm1$.

The dynamic resistance is defined to be
\begin{equation}
\label{21}
R_D=\frac{d\bar{V}}{dI},\,\, r_D\equiv z_0 =\frac{d\bar{v}}{di},
\end{equation}
where $\bar{V}$ denotes the time-average over one cycle. In this approximation ($\beta=0$)
\begin{equation}
\label{22}
\bar{v}=\frac{2\pi}{\tau_{J0}}=\omega_{J0},
\end{equation}
so, from \eqref{20},
\begin{equation}
\label{23}
r_{D0}=\frac{i}{\omega_{J0}}.
\end{equation}

 Equation \eqref{15} gives
\begin{eqnarray}
\label{24}
\langle V_{n0}(0)V_{0n}^*(0)\rangle _{\hat{t}}
&=&\sum_{k,k'}\exp[-j\hat{t}\Omega_J(k-k')]
\nonumber \\ &\times&
Z_kZ^*_{k'}\langle I_{nk}(0)I^*_{nk'}(0)\rangle
\end{eqnarray}
where we have made explicit the $\hat{t}$ dependence in \eqref{14} inherited from \eqref{9}.
We now average \eqref{24} over $\hat{t}$   
through one period $T_J$  and use \eqref{2}  to get 
\begin{eqnarray}
\label{25}
S_V(0)&=&\langle V_{n0}(0)V^*_{n0}(0)\rangle
\nonumber \\
&\equiv&\dfrac{1}{T_J}\int_0^{T_J}
d\hat{t}\langle V_{n0}(0)V^*_{n0}(0)\rangle_{\hat{t}}
\nonumber \\
&=&\sum_k |Z_k|^2S_I(k\Omega_J).
\end{eqnarray}
Since $Z^*_k=Z_{-k}$, the right hand side of this equation automatically contains the
symmetrized products of currents, as in equation \eqref{1}.
Finally using \eqref{1} with $\Omega=k\Omega_J$, \eqref{25} gives 
the required prediction for the low frequency voltage noise $S_V(0)$. 
\begin{widetext}
\begin{eqnarray}
\label{26}
S_V(0)
=
2 \sum_k |Z_k|^2\left(\frac{\hbar k\Omega_J}{R}\right)\coth
\left[\frac{\hbar k\Omega_J}{2T}\right] 
=
4 eR^2I_0 \left[\dfrac{|z_0|^2}{p}+
\sum_{k\geq1}k|z_k|^2\omega_J \coth(p k\omega_J)\right], 
\end{eqnarray}
\end{widetext}
where we have used \eqref{17}  and $p=eRI_0/T$ (and $p$ is close to $1$ in the experiment. \cite{koch}).
     
For the special case $C\simeq0$, we have
\begin{equation}
\label{27}
Z_0=\frac{RI}{\sqrt{I^2-I_0^2|}}, \;\; |Z_{\pm1}|=\frac{RI_0}{2\sqrt{I^2-I_0^2}},
\end{equation}
and all other $Z_k=0$.
Then  \eqref{26} gives
\begin{eqnarray}
\label{28}
S_V(0)&=&\frac{4TRI^2}{I^2-I_0^2}+\frac{eR^2I_0^2}{\sqrt{I^2-I_0^2}}
\nonumber \\ &\times& 
\coth\left[\frac{eR\sqrt{I^2-I_0^2}}{2T}\right].
\end{eqnarray}
Using  \eqref{17}, \eqref{20}, \eqref{21} and \eqref{22}, 
this may alternatively be written
\begin{equation}
\label{29}
\frac{S_V(0)}{R_D^2} = \frac{4T}{R}+
\frac{2 e\bar{V}I_0^2}{RI^2}\coth\left[\frac{e\bar{V}}{T}\right].
\end{equation}
It has been suggested \cite{koch} that in this form, if observed values of $\bar{V}$ and $R_D$
are used, it may also be a good approximation for non-zero  $\beta$ (defined in \eqref{17}).
{
One purpose of this paper is to test this approximation. }
We note from \eqref{26} that the quantity $S_V(0)/R_D^2$ in \eqref{29}
depends upon the magnitudes of the ratios {
$(Z_k/R_D)^2$. } 
\section{Perturbation theory for small capacitance}\label{sec3}

This section is about the calculation of the
Josephson frequency; it is not concerned with noise, so all variables
are classical ones.

Since $\beta$ in equation \eqref{18}
 has modest values 
 in the experiments \cite{koch}, one might expect
perturbation theory in $\beta$ to be useful. This is especially so as the series turns out to
be an expansion in $\beta^2$. This property follows because  \eqref{18} is invariant under
\begin{equation}
\label{30}
\tau\rightarrow -\tau,\;\; \beta\rightarrow -\beta,\;\; \theta\rightarrow \pi-\theta,\;\;i\rightarrow +i.
\end{equation}

To construct the perturbation series, we expand
\begin{equation}
\label{31}
\theta=\theta_0 +\beta\theta_1 + \beta^2 \theta_2 + \dots,
\end{equation}
and\begin{equation}
\label{32}
\tau_J=\tau_{J0} +\beta \tau_{J1}+ \beta^2 \tau_{J2} + \dots,
\end{equation}
where $\tau=2\pi/\omega_J$ is the Josephson period and $\tau_{J0}$ is given by \eqref{20}.
Inserting \eqref{31} into \eqref{18},
\begin{equation}
\label{33}
\theta'_1+(\cos\theta_0)\theta_1=-\theta''_0,
\end{equation}
\begin{equation}
\label{34}
\theta'_2+(\cos\theta_0)\theta_2=-\theta''_1+\frac{\theta_1^2}{2}\sin\theta_0 ,
\end{equation}
where prime denotes differentiation with respect to $\tau$.

Equation \eqref{33} gives, using \eqref{19},
\begin{equation}
\label{35}
[i +\sin(\omega_{J0}\tau)\theta_1]'=\frac{\omega_{J0}^3\cos(\omega_{J0}\tau)}{i+\sin(\omega_{J0}\tau)}.
\end{equation}
Hence
\begin{equation}
\label{36}
\theta_1=\frac{\omega_{J0}^2
\ln\left[1+\sin(\omega_{J0}\tau)/i\right]}{i+\sin(\omega_{J0}\tau)}
\end{equation}
where we have chosen the initial value $\theta(0)=1/i$. This solution has the same frequency
$\omega_{J0}$ as  $\theta_0$; so $\tau_1=0$ in \eqref{32}. This is in accordance with our remark
above that the expansion is  series in $\beta^2$.

From the definition of $T$, for any $\tau$, to the requisite order,
\begin{eqnarray}
\label{37}
2\pi&=&\theta(\tau+\tau_J)-\theta(\tau)
\nonumber \\
&=&\theta_0(\tau+\tau_J)-\theta_0(\tau)
\nonumber \\
&+&\beta^2[\theta_2(\tau+\tau_{J0})-\theta(\tau)],
\end{eqnarray}
giving 
\begin{equation}
\label{38}
\tau_{J2}=-[\theta_2(\tau+\tau_{J0})-\theta_2(\tau)]/\theta'_{J0}(\tau).
\end{equation}
(To be consistent, the right hand side of this equation must turn out to be independent of $\tau$.)

Then, from \eqref{34},  \eqref{36} and \eqref{38}, and using \eqref{19}, 
\begin{eqnarray}
\label{39}
\tau_{J2}&=&\dfrac{1}{{\omega_{J0}}^2}\int_{\tau}^{\tau+\tau_{J0}}
d\tau'[i+\sin(\omega_{J0}\tau')]
\nonumber \\
&\times&[\theta''_1(\tau')
-\dfrac{1}{2}(\sin\theta_0){\theta_1(\tau')}^{2}]
\nonumber \\
&=&-\omega_{J0}^2\int_{\tau}^{\tau+\tau_{J0}}d\tau'
\nonumber \\ &\times& 
\left[\sin(\omega_{J0}\tau')\ln\left(1+\dfrac{\sin(\omega_{J0}\tau')}{i}\right)
\right . \nonumber \\
&+&\left . \dfrac{1+i\sin(\omega_{J0}\tau')}
{\left(i+\sin(\omega_{J0}\tau')\right)^2}
\right . \nonumber \\
&\times& \left.
\ln^2\left(1+
\dfrac{\sin(\omega_{J0}\tau')}{i}\right)\right],
\end{eqnarray}
where we have integrated by parts the $\theta''_1$ term in the first line.
It turns out that all the integrals arising in \eqref{39} which contain a power of $\ln(1+\sin(\omega_{J0}\tau)/i)$ can be done by integration by parts:
\begin{eqnarray}
\label{40}
\tau_{J2}&=&-\omega_{J0}\int_{\tau}^{\tau+\tau_{J0}}d\tau'
\nonumber \\ & \times & 
\left[\frac{\omega_{J0}\cos^2(\omega_{J0}\tau')}{(i+\sin(\omega_{J0}\tau'))^2}-\frac{dH}{d\tau'}\right],
\end{eqnarray}
Where
\begin{eqnarray}
\label{41}
H(\tau)&=&\frac{\cos(\omega_{J0}\tau)}{i+\sin(\omega_{J0}\tau)}
\left[\frac{1}{2}
\ln^2\left(1+\frac{\sin\omega_{J0}\tau}{i}\right)
\right. \nonumber \\ 
&-&\left.\ln\left(1+\frac{\sin(\omega_{J0}\tau}{i}\right)+1\right].
\end{eqnarray}
It is now a simple matter to  complete the integration and find
\begin{eqnarray}
\label{42}
\tau_{J2}&=&
-\omega_{J0}^2\int_\tau^{\tau+\tau_{J0}} d\tau'
\frac{\cos^2(\omega_{J0}\tau')}{(i+\sin(\omega_{J0}\tau'))^2}
\nonumber \\
&=&-2\pi(i-\omega_{J0}).
\end{eqnarray}

Thus the Josephson period to second order is\begin{equation}
\label{43}
\tau_J=\tau_{J0}[1-\beta^2\omega_{J0}(i-\omega_{J0})],
\end{equation}
and the frequency is\begin{equation}
\label{44}
\omega_J=\omega_{J0}[(1+\beta^2\omega_{J0}(i-\omega_{J0})].
\end{equation}
From  \eqref{22}, in this approximation the dynamical resistance is
\begin{equation}
\label{45}
r_D=z_0=\frac{i}{\omega_{J0}}+\beta^2(i-\omega_{J0})(2i-\omega_{J0}).
\end{equation}

Figures \ref{fig1} and \ref{fig2} show a comparison of \eqref{44} and
\eqref{45} with numerical results.

\section{Perturbation theory for impedance of higher harmonics}\label{sec4}

In this  section, we use perturbation theory in $\beta$ to study the impedances $z_k$.
Because $z_1$ has a contribution (see \eqref{27}) to zeroth order  in $\beta$,
\[
|z_1|^2=|z_{10}+\beta z_{11}+\beta^2 z_{12}+\dots|^2
\]
 and we need the second order term because of its interference with the zeroth order one;
 so first order perturbation theory on its own has no physical significance.
 We have found second order perturbation theory prohibitively complicated.
 But for $k>1$, there is  no zeroth order contribution and so first order perturbation theory
 is relevant by itself, and this we now study.

{
Since our aim is to calculate the impedances $z_k$ (which are classical quantities), it is sufficient
to introduce, instead of the quantum noise $i_n$, a classical driving force at the Josephson
frequency $\omega_J$ into \eqref{18}. To our approximation, it is sufficient to
use Eq. (2.4). Thus we use the equation
}
\begin{equation}
\label{46}
\beta\theta''+\theta'+\sin\theta=i+a_k\exp(jk\omega_{J0}\tau)
\end{equation}
where $a_k$ is complex and infinitesimal. We expand
\begin{equation}
\label{47}
\theta=\theta_0+a_k\theta_{a_k}+b\theta_1+ba_k\theta_{1a_k}+\dots
\end{equation}
Then
\begin{equation}
\label{48}
\theta'_{a_k}+\theta_{a_k}\cos\theta_0=\exp(ik\omega_{J0})
\end{equation}
so
 \begin{eqnarray}
\label{49}
&[(i+\sin(\omega_{J0}\tau))]\theta_{a_k}=F
\nonumber \\
&\equiv \int_{\tau_0}^{\tau+\tau_0} d\tau'[(i+\sin(\omega_{J0}\tau')]\exp(jk\omega_{J0}\tau').
\end{eqnarray}
$\theta_1$ is given in \eqref{36}.

To the next order,
\begin{equation}
\label{50}
\theta'_{1a_k}+(\cos\theta_0)\theta_{1a_k}=(\sin\theta_0)\theta_{a_k}\theta_1-\theta''_{a_k}.
\end{equation}
So, using \eqref{19} and \eqref{50},
\begin{eqnarray}
\label{51}
&\dfrac{d}{d\tau}[(i+\sin(\omega_{J0}t))\theta_{1a_k}]=
\nonumber \\
&\omega_{J0}^2\dfrac{1+i\sin(\omega_{J0}\tau)}{i+\sin(\omega_{J0}t\tau)}
\ln\left[1+\dfrac{\sin(\omega_{J0}\tau)}{i}\right]\theta_{a_k}
\nonumber \\
&-\left[i+\sin(\omega_{J0}\tau)\right]\theta''_{a_k}.
\end{eqnarray}

We want to determine the period of $\theta$ in \eqref{47}. Let this period be
\begin{equation}
\label{52}
\tau_J=\tau_{J0}+ba_k\tau_{1a_k}+\dots
\end{equation}
then
\begin{eqnarray}
\label{53}
2\pi&=&\theta_0(\tau+\tau_{J0}+ba_k\tau_{1a_k})-\theta_0(\tau)
\nonumber \\
&+&ba_k[\theta_{1a_k}(\tau+\tau_{J0})-\theta_{1a_k}(\tau)]
\end{eqnarray}
where $\tau$ is  arbitrary, so
\begin{equation}
\label{54}
\tau_{1a_k}\theta_0'(\tau)=-[\theta_{1a_k}(\tau+\tau_{J0})-\theta_{1a_k}(\tau)].
\end{equation}

It follows from \eqref{4} and \eqref{51} that
\begin{eqnarray}
\label{55}
\omega_{J0}^2\tau_{1a_k}&=&
-\displaystyle{\int_{\tau}^{\tau+\tau_{J0}}d\tau'}
\left[\omega_{J0}^2\dfrac{1+i\sin(\omega_{J0}\tau')}{(i+\sin(\omega_{J0}\tau'))^2}
\right . \nonumber \\ &\times&  \left .
\ln[1+\dfrac{\sin\omega_{J0}\tau}{i}]\theta_{a_k}(\tau')
\right . \nonumber \\ & - &  \left .
(i+\sin(\omega_{J0}\tau'))\theta''_{a_k}(\tau')\right].
\end{eqnarray}
Integrating by parts and using the periodicity of the integrand, the second term in the square
bracket in \eqref{55} is converted to
\begin{equation}
\label{56}
\omega_{J0}^2\sin(\omega_{J0}\tau')\theta_{1a_k}.
\end{equation}
Then, using \eqref{49},  \eqref{55} gives
\begin{eqnarray}
\label{57}
\tau_{1a_k}
&=&
 -\dfrac{1}{\omega_{J0}}\displaystyle{\int_{\tau}^{\tau+\tau_{J0}} dt'}
\left\{F'(\tau')\dfrac{\cos(\omega_{J0}\tau')}{\sin(\omega_{J0}\tau')}
\right . \nonumber \\ &\times & \left .
\left[1+\ln(1+\dfrac{\sin(\omega_{J0}\tau')}{i})\right] 
-G'(\tau')\right\}
\end{eqnarray}
where
\begin{eqnarray}
\label{58}
G(\tau)&=&\frac{\cos(\omega_{J0}\tau)}{i+\sin(\omega_{J0}\tau)}F(\tau)
\nonumber \\ &\times&
\left[1+\ln\left(1+\frac{\sin(\omega_{J0}\tau)}{i}\right)\right]
\end{eqnarray}
Again, because of the periodicity, the contribution from the $G'$ term in \eqref{57} is zero.
Finally, inserting the value of $F'$ from \eqref{49} and omitting an integral of a differential of a  periodic function,
we get
\begin{eqnarray}
\label{59}
\tau_{1a_k}&=&-\frac{1}{\omega_{J0}}\int_{\tau}^{\tau+\tau_{J0}}
d\tau'\cos(\omega_{J0}t')
\nonumber \\ &\times&\ln\left[1+\frac{\sin(\omega_{J0}\tau')}{i}\right]
\exp(jk\omega_{J0}\tau').
\end{eqnarray}

 Because the integrand in \eqref{59} is periodic, we may replace the limits of integration by $0$ and $\tau_{J0}$
Integrating by parts again, it gives
\begin{eqnarray}
\label{60}
\tau_{1a_k}&=&\dfrac{1}{2\omega_{J0}}
\int_0^{\tau_{J0}}d\tau'\frac{\cos(\omega_{J0}\tau')}{i+\sin(\omega_{J0}\tau')}
\nonumber \\ &\times&
\left[\frac{\exp(j(k+1)\omega_{J0}t\tau')}{k+1}
+\frac{\exp(j(k-1)\omega_{J0}\tau')}{k-1}\right].
\nonumber \\ &&
\end{eqnarray}
The integral may now be done by the substitution $\zeta=\exp(j\omega_{J0}\tau')$ and integrating
round the unit circle. There are poles at $\zeta=-ju, -ju^{-1}$ where\begin{equation}
\label{61}
u=i+\omega_{J0},\;\;\;\; u^{-1}=i-\omega_{J0},
\end{equation}
and  $-ju^{-1}$ lies within the unit circle (taking $i>1$ for the moment).
Remembering that
\begin{equation}
\label{62}
z_k=\beta\omega_{1a_k}=-\beta\omega_{J0}^2 \frac{\tau_{1a_k}}{2\pi},
\end{equation}
the result is
\begin{equation}
\label{63}
z_k=(-j)^{k+1}\frac{\beta}{2}\left[\frac{u^{-k+1}}{k-1}-\frac{u^{-k-1}}{k+1}\right].
\end{equation}
(The restriction $i\neq 1$ may now be relaxed.)

For low enough temperature, or high enough current $I$, 
{
when $p$ in (2.10) is considerably greater than 1,
}
equation \eqref{26} for the voltage noise approximately contains the sum 
\begin{equation}
\label{64}
\sum_{k>1}k|z_k|^2
\end{equation}
which, in the small $\beta$ approximation \eqref{63}, may be evaluated to give
\begin{equation}
\label{65}
\beta^2\left[\frac{5}{16}-
\frac{\omega_{J0}^2}{4(i+\omega_{J0})^2}
+\omega_{J0}^2\ln\left(\frac{2\omega_{J0}}{i+\omega_{J0}}\right)\right].
\end{equation}
{
In the experiment \cite{koch}, 
$k$ is of order 1, but still \eqref{64} } 
is a lower bound to the contributions from $k>1$ (since the $\coth$ factors are greater than 1).

\section{Perturbation theory for high current}\label{sec5}
In this section, we develop an expansion in powers of $1/i=I_0/I$. This is algebraically simpler
than the expansion in $\beta$, but is relevant only to a small range of $I$ in the experiments.

Because of the invariance of \eqref{18} under $i\rightarrow -i, \theta\rightarrow -\theta$, the
expansion is in powers of $1/i^2$.

Define $s=i\tau$ and a function $\vartheta(s)=\theta(\tau)$. Then \eqref{46} becomes
\begin{equation}
\label{66}
(i\beta)\vartheta''+\vartheta'+\frac{1}{i}\sin\vartheta=
1+\frac{a_k}{i}\exp(jks),
\end{equation}
and we will be concerned only with $k=1$ or $2$ in this section. In general the frequency
of the last term in \eqref{66}   ought to be $\omega_J$, but to our order it is sufficient to use the
zeroth approximation $\omega_{J0}\approx i$.
{
Since the driving force is classical, $\theta$ is
also classical in this section and in section \ref{sec6} }.

Expand
\begin{eqnarray}
\label{67}
\vartheta&=&\vartheta_0+
\frac{1}{i}\vartheta_1+
\frac{1}{i^2}\vartheta_2
\nonumber \\ &+&
\frac{a_1}{i}\vartheta_{1a_1}+
\frac{a_1}{i^2}\vartheta_{2a_1}+
\frac{a_2}{i}\vartheta_{1a_2}+\dots
\end{eqnarray}
with similar expansions 
of $\tau_J$ and $\omega_J$.  
(We will find that $\tau_{na_k}=0$ for $n< k+1$.) 
We take $\vartheta_0=s$, being the solution
without a transient decreasing  exponential. Then
\begin{equation}
\label{68}
(i\beta)\vartheta''_I+\vartheta'_1=-\sin\vartheta_0=-\sin s
\end{equation}
\begin{equation}
\label{69}
(i\beta)\vartheta''_{1a_1}+\vartheta'_{1a_1}=\exp(js),
\end{equation}
\begin{equation}
\label{70}
(i\beta)\vartheta_2''+\vartheta_2'=-(\cos\vartheta_0)\vartheta_1=-(\cos s)\vartheta_1,
\end{equation}
\begin{equation}
\label{71}
(i\beta)\vartheta''_{2a_2}+\vartheta'_{2a_2}=-(\cos\vartheta_0)\vartheta_{1a_1}=-(\cos s)\vartheta_{1a_1}.
\end{equation}

The solution of \eqref{68} with no transient, and a convenient initial value, is
\begin{equation}
\label{72}
\vartheta_{1}=\frac{1}{2}\left[\frac{\exp(js)}{1+jg}+\frac{\exp(-js)}{1-jg}\right]
\end{equation}
where\begin{equation}
\label{73}
g=i\beta.
\end{equation}
Since \eqref{72} is periodic with the same period 
as $\vartheta_0$, the period is not changed to
first order, as expected.

The solution of \eqref{69} is
\begin{equation}
\label{74}
\vartheta_{1a_1}=\frac{-j\exp(js)}{1+jg}.
\end{equation}
Again, the period is unaltered, and so there is 
no correction to $\bar{v}$ or to $z_0$ at this order.

Inserting \eqref{72} into \eqref{70}, 
\begin{eqnarray}
\label{75}
(i\beta)\vartheta''_2+\vartheta_2&=&-\dfrac{1}{4(1+g^2)}
\nonumber \\ &\times &
\left[2+(1-jg)\exp(2js)
\right . \nonumber \\ 
&+&\left .(1+jg)\exp(-2js)\right].
\end{eqnarray}
so
\begin{equation}
\label{76}
\vartheta_2(s+2\pi)-\vartheta_2(s)=-\frac{\pi}{1+g^2}.
\end{equation}
It follows, in a similar manner to \eqref{53}, the period at this order is 
\begin{equation}
\label{77}
\tau_{J0}+\frac{1}{i^2}\tau_{J2}=\frac{2\pi}{i}\left[1+\frac{1}{2i^2(1+g^2)}\right],
\end{equation}
and
\begin{eqnarray}
\label{78}
\bar{v}&=&\omega_J=i\left[1-\frac{1}{2i^2(1+g^2)}\right], 
\nonumber \\
r_D&=&\frac{d\omega_{J}}{di}=1+\frac{1+3g^2}{2i^2(1+g^2)^2},
\end{eqnarray}
to this order.
(For $\beta=0$, this is consistent with \eqref{20}).

Inserting \eqref{74} into \eqref{71}, we find that
\begin{equation}
\label{79}
\vartheta_{2a_2}=-\frac{\exp(2js)}{[2(j-g)(2j-4g)]} -\frac{s}{[2(j-g)]},
\end{equation}
hence
\begin{equation}
\label{80}
\vartheta_{2a_2}(s+2\pi)-\vartheta_{a2}(s)=-\pi \frac{1}{(j-g)}.
\end{equation}
Therefore
\begin{equation}
\label{81}
\tau_{2a_1}=\pi\frac{i}{(j-g)},
\end{equation}
and
\begin{equation}
\label{82}
z_1=-\frac{1}{2\pi i^2}\tau_{2a_1}=-\frac{1}{2i(j-g)},\;\;\;|z_1|=\frac{1}{2i\sqrt{1+g^2}}.
\end{equation}
Comparisons of \eqref{78} and \eqref{82} with numerical data 
are shown in Figs. \ref{fig1}, \ref{fig2} and \ref{fig3}.  The perturbation theory
is in reasonable agreement for $i>2.0$.

 We can use the expansion in $1/i$ for $k=2$, but we must go to one order higher in $1/i$
to find $z_2$.  We use the $a_2$ terms in \eqref{66} and \eqref{67}.
The equations are
\begin{eqnarray}
\label{83}
g\vartheta''_{1a_2}+\vartheta'_{1a_2}&=&\exp(2js),
\nonumber \\
g\vartheta''_{2a_2}+\vartheta'_{2a_2}&=&-(\cos s)\vartheta_{1a_2}, \\
g\vartheta''_{3a_2}+\vartheta'_{3a_2}&=&
-(\cos s)\vartheta_{2a_2}+(\sin s)\vartheta_1\vartheta_{1a_2}.
\nonumber
\end{eqnarray}
Then
\begin{eqnarray}
\label{84}
\vartheta_{1a_2}&=&\frac{\exp(2js)}{2j-4g}
\nonumber \\
\vartheta_{2a_2}&=&-\frac{\exp(js)}{[2(j-g)(2j-4g)]}
\nonumber \\
&-&\frac{\exp(3js)}{[2(9g-3j)(4g-2j)]},
\end{eqnarray}
where the second term will turn out to be irrelevant,

Putting \eqref{84} into the last of \eqref{83} and using \eqref{72},
\begin{eqnarray}
\label{85}
g\vartheta''_{3a_2}+\vartheta'_{3a_2}&=&\frac{1}{[4(j-g)(2j-4g)]}
\nonumber \\
&-&\frac{1}{[4(2j-4g)(j+g)]}+\dots
\end{eqnarray}
where the omitted terms are periodic and irrelevant.
Hence
\begin{equation}
\label{86}
\vartheta_{3a_2}(s+2\pi)-\vartheta_{3a_2}(s)=\frac{2\pi g}{[4(j-2g)(g^2+1)]}.
\end{equation}
Thus we find that, to this order,
 \begin{equation}
\label{87}
|z_2|=\frac{|g|}{[4i^2(g^2+1)\sqrt{(4g^2+1)}]}.
\end{equation}
As expected, this is zero if $\beta=0$ (since $g=i\beta$). 

The above formulae \eqref{78},  \eqref{82} and \eqref{87} 
are compared with numerical results in Figs. \ref{fig1}, \ref{fig2}, \ref{fig3} and 
\ref{fig5}, respectively.
{
As expected, the perturbation theory is not accurate for the smaller values of $i$, but
it agrees quite well with the numerical results for $i>2$.
}

\section{Numerical results}\label{sec6}
We start with equation \eqref{18}, including real forcing terms:
\begin{equation}
\label{88a}
\beta\theta''+\theta'+\sin\theta=i+a_s \sin(\omega'\tau)+a_c\cos(\omega'\tau),
\end{equation}
where $a_s$ and $a_c$ are small (in practise we have verified that, in the range  $10^{-4} \le a_s$, $a_c \le 10^{-3}$, linearity holds within the numerical precision of the calculations), and $\omega'$ is adjusted so that\begin{equation}
\label{89}
\omega_J(\omega', a_s, a_c)=\omega',
\end{equation}
$\omega_J$ being the frequency of the solutions of \eqref{88a} (after the decay of transients).

\begin{figure}[t!] 
\centering
\includegraphics[scale=0.65]{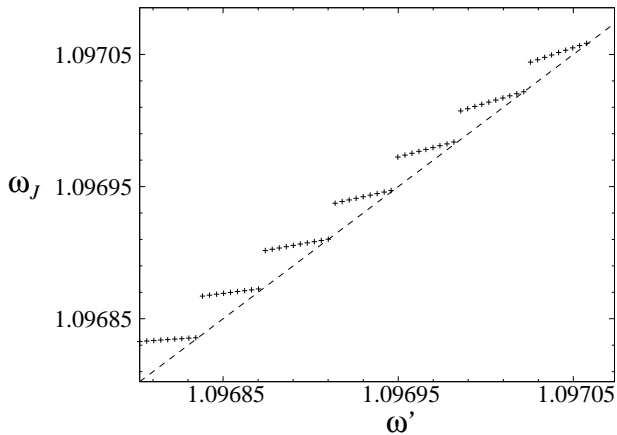}
\caption{Graphical representation of
the equation $\omega_J(\omega', a_s, 0)=\omega'$.
The crosses are the numerical results for  $\omega_J(\omega', a_s, 0)$
where $a_s=10^{-4}$, $2\times 10^{-4}$, $\dots$ and $7\times 10^{-4}$.
In this example $\beta=0.45$ and $i=1.4142$.}\label{fig0}
\end{figure}

\begin{figure}[t!] 
\centering
\includegraphics[scale=0.65]{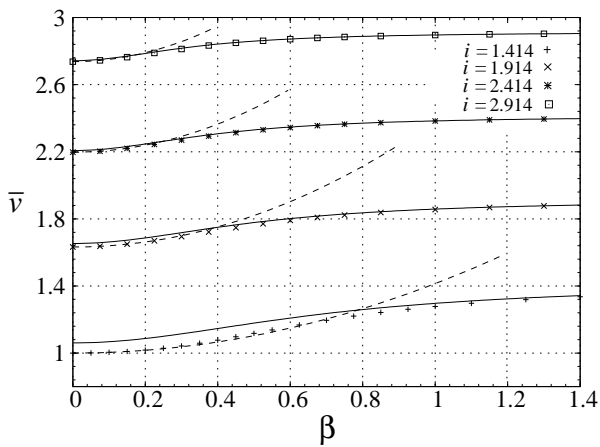}
\caption{The dimensionless average voltage as a function of $\beta$ and
four values of $i$. The continuous and broken lines are the 
perturbative results in powers of $1/i$ and $\beta$ respectively.} \label{fig1}
\end{figure}

\begin{figure}[t!] 
\centering
\includegraphics[scale=0.65]{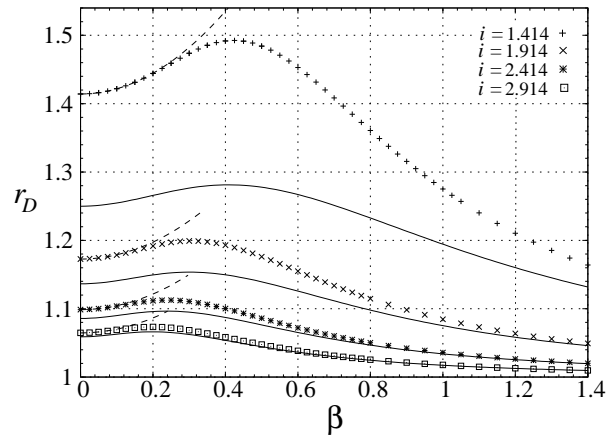}
\caption{The dimensionless dynamical resistance as a function of $\beta$ and
four values of $i$. The continuous and broken lines are 
the perturbative results in powers of $1/i$ and $\beta$ respectively.}\label{fig2} 
\end{figure}

\begin{table}[t!] 
\centering
\begin{tabular}{|c|c|r|}
\hline 
$\beta$ & $\frac{\Delta\bar v}{\bar v}\times 100$  & $\frac{\Delta r_D}{r_D}\times 100$     \\ 
\hline\hline
                    0.00  &       6.066  &      11.611 \\
                    0.30  &       6.923  &      13.380 \\
                    0.60  &       4.371  &      12.790 \\
                    0.90  &       1.921  &       7.760 \\
                    1.20  &       0.854  &       4.168 \\
                    1.50  &       0.415  &       2.264 \\
                    1.80  &       0.221  &       1.289 \\
                    2.10  &       0.126  &       0.772 \\
                    2.40  &       0.077  &       0.485 \\
                    2.70  &       0.049  &       0.318 \\
                    3.00  &       0.033  &       0.216 \\
                    3.30  &       0.023  &       0.152 \\
                    3.60  &       0.016  &       0.109 \\
\hline
\end{tabular}
\caption{Differences between the numerical and 
the $1/i$ perturbation theory for both $\bar v$ and $r_D$, for $i=1.41$.}
\label{tab1}
\end{table}

\begin{figure}[t!] 
\centering
\includegraphics[scale=0.65]{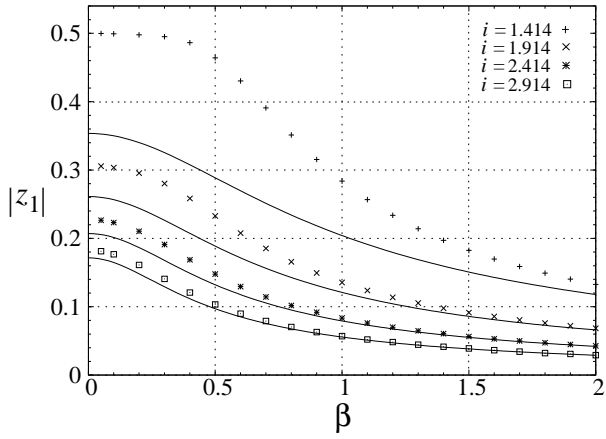}
\caption{The impedance associated with the first harmonic as a function of $\beta$ and
four values of $i$. The full lines represent the $1/i$ perturbative result.}\label{fig3}
\end{figure}

\begin{figure}[t!] 
\centering
\includegraphics[scale=0.65]{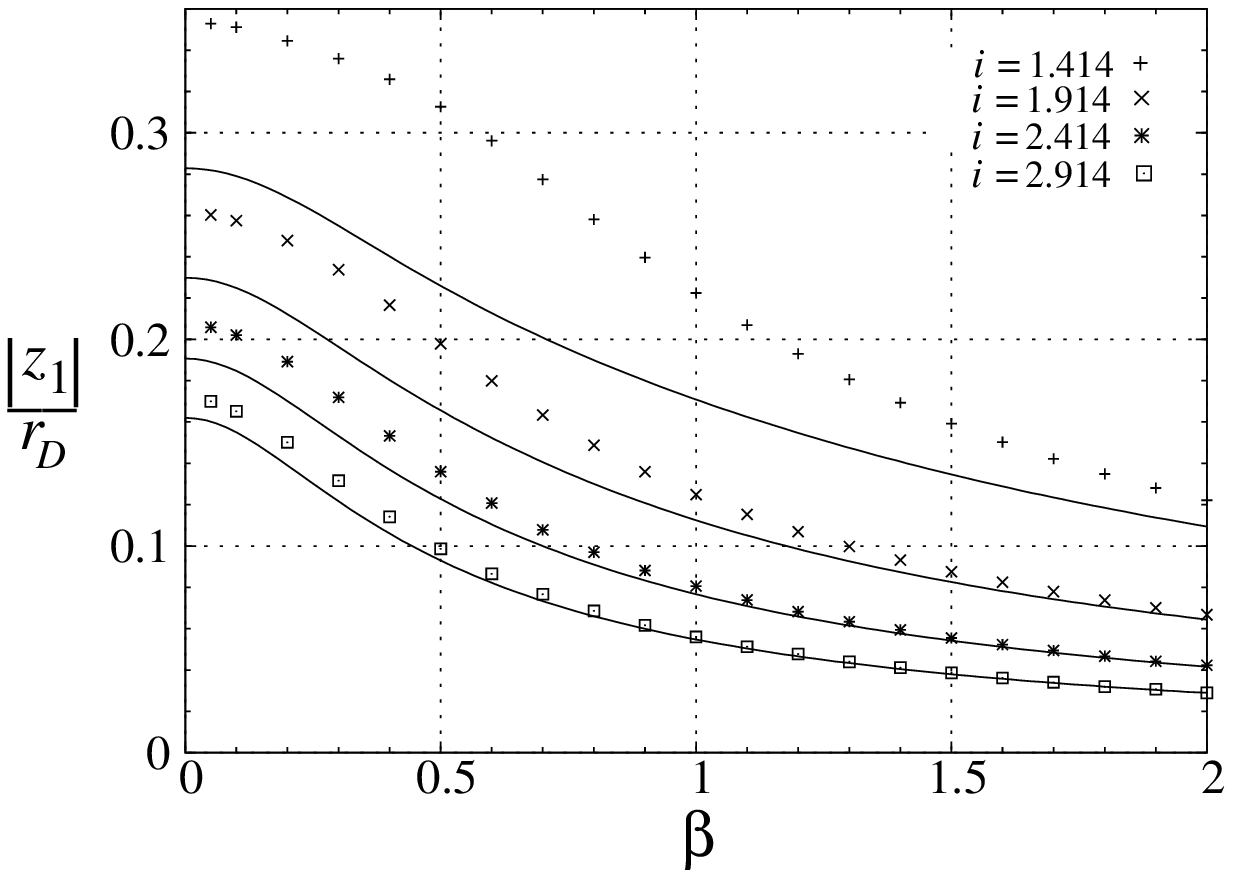}
\caption{Same as in figure \ref{fig3} but divided by $r_D$.}\label{fig4}
\end{figure}

\begin{figure}[t!] 
\centering
\includegraphics[scale=0.65]{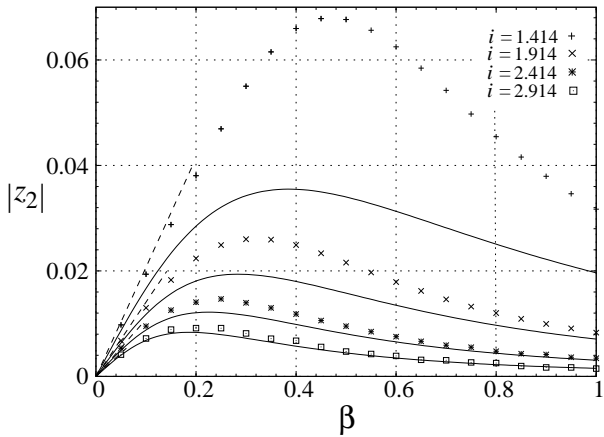}
\caption{The impedance associated with the second harmonic as a function of $\beta$ and
four values of $i$. The full lines are the $1/i$  perturbative results and
the dashed lines are the order $\beta$
perturbative result given by \eqref{63}.} \label{fig5}
\end{figure}

\begin{figure}[t!] 
\centering
\includegraphics[scale=0.65]{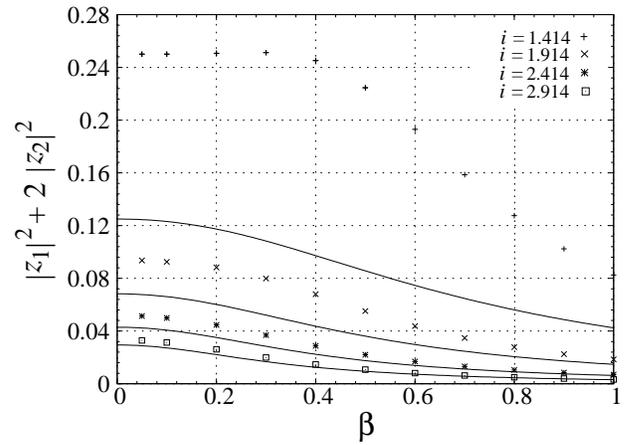}
\caption{$|z_1|^2 + 2 |z_2|^2$ as a function of $\beta$ and four 
values of $i$.  The full lines are the $1/i$  perturbative results.}\label{fig6}
\end{figure}

\begin{figure}[t!] 
\centering
\includegraphics[scale=0.65]{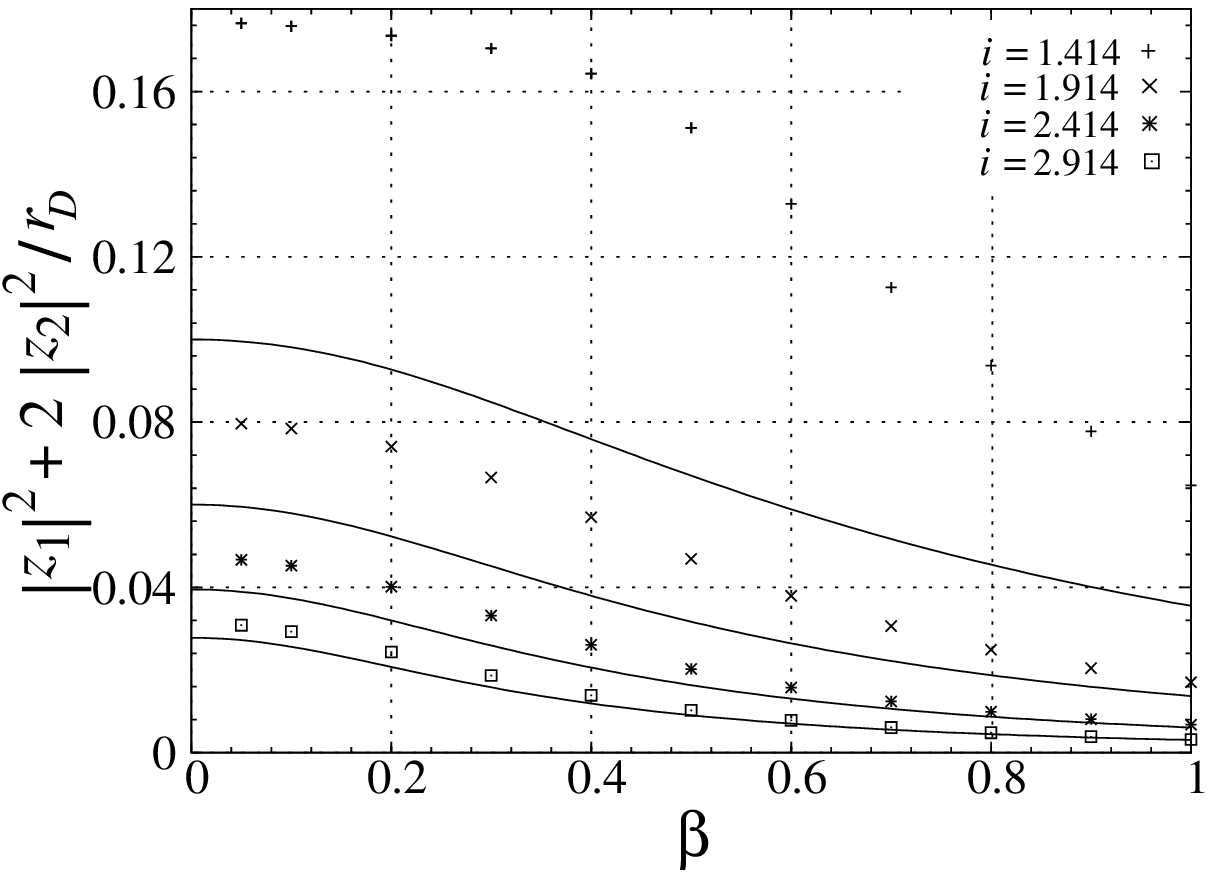}
\caption{Same as in figure \ref{fig6} but divided by $r_D$.}\label{fig7}
\end{figure}

The numerical solution of Eq. \eqref{88a} was investigated using several 
initial conditions for both $i<1$ and $i>1$. Our main interest here are the cases
when $i>1$. In these cases, we start with arbitrary initial conditions 
at $\tau=0$, and begin measurements on the
solution at $\tau=\tau_0$ where $\tau_0$ is about $100$  (when
transients have become negligibly small, for all relevant choices of
$\beta$, $i$ and $\omega'$). 
Then, after $N=10^3$ cycles, 
the condition $\theta(\tau_1)-\theta(\tau_0) = 2\pi N$ yields the period 
$\tau_J=(\tau_1-\tau_0)/N$ (the size of the time step employed in the numerical
code was $\Delta\tau =10^{-4}$, between
$\tau=0$ and $\tau=\tau_0$, and $\Delta\tau =10^{-6}$ after the transients).
At the solutions of \eqref{89}, but not otherwise, the results for 
$\omega_J=2\pi/\tau_J$ and $|z|$  are stable against changes in the initial
conditions and in $\tau_0\simeq 100$ and $N\simeq 10^3$. 

The next step in the numerical 
procedure consists in solving \eqref{89}. In the figure \ref{fig0} we illustrate this
procedure with an example of the intersections 
of $\omega_J(\omega', a_s, 0)$ with the diagonal line for $\beta=0.45$, $i=1.4142$
and a series of values of $a_s$. Although figure \ref{fig0} 
only shows a few points in the vicinity of the diagonal line, the 
numerical code employed 40 times more points, in the same range shown
in figure \ref{fig0}, in order
to reach the required precision in the determination of the impedances.
In this range of variation of $a_s$,
the point of intersection increases linearly with $a_s$. 

In general the numerical procedure solves \eqref{89} and yields the result
\begin{equation}
\label{88}
\omega_J(a_s,a_c)=\omega_J(0,0)+a_sz_s+a_cz_c+\dots
\end{equation}
which gives the Josephson frequency and the two real impedances (for $k=1$).
(We also have obtained the nonlinear terms in \eqref{88} which are not 
important in the present analysis).
Using this approach, we were able to perform a detailed numerical calculation of
the quantities 
$\omega_J$,  $r_D$, $|z_1|=\sqrt{z_s^2+z_c^2}$ and $|z_1|/r_D$. 
We also have investigated the higher harmonics ($k=2, 3, \dots$) and computed 
the corresponding quantities such as $|z_2|$, $|z_2|/r_D$ and $|z_1|^2+2|z_2|^2$. 
Our main interest is to investigate how significant is the dependence on $\beta$
for several values of $i$.

Let us start with the results for $\bar{v}=\omega_J(0,0)$ 
and $r_D$ (see Eq. \eqref{78}).  These are shown in figures \ref{fig1} and \ref{fig2}
as functions of $\beta$ and four values of $i$. From these figures one can see 
what are the numerical values of $\beta$ and $i$ such that the perturbative results,
obtained in the previous sections, can be trusted. For instance,
the perturbative increase with $\beta$ only persists up to about  
$\beta=0.3$, for $i=1.4$; after that $r_D$ decreases.
Although the numerical computation has been performed up to $\beta=4$, 
here we are focusing on the interval which may be more realistic 
from the phenomenological point of view. Besides, larger values of 
$\beta$ are very well described by the $1/i$ perturbation theory even for $i=1.4$. 
For instance, we have found that, for $i=1.4$, the difference between 
the numerical and the $1/i$ perturbative result for $r_D$ is smaller than 1\% for 
$\beta>2.0$ (see table \ref{tab1}).

Let us now consider the results for the impedances associated with the forcing
terms. The numerical results for 
$|z_1|=\sqrt{z_s^2+z_c^2}$ and $|z_1|/r_D$ are shown in figures
\ref{fig3} and \ref{fig4}, respectively. 
Similarly to the case of the dynamical resistance 
the curves for the $1/i$  perturbative result (full lines) 
underestimate the exact numerical points by an amount which
becomes negligible when $i$ increases. Also, 
the perturbative result becomes indistinguishable from the exact result
for $\beta$ sufficiently large.

The results for the impedances associated with second harmonic are
shown in figure \ref{fig5}. The validity of the perturbative
results for small $\beta$ or large $i$ can be seem in the figure.
As in the previous cases the curves for the $1/i$  
perturbative result become very close to the exact numerical result
for $\beta$ sufficiently large. 
Finally, in figures \ref{fig6} and \ref{fig7}
we shown the results for $|z_1|^2 + 2 |z_2|^2$ and $(|z_1|^2 + 2 |z_2|^2)/r_D$.
These quantities are  relevant for the calculation of the low frequency
voltage noise  (according to the Eqs. \eqref{26} and \eqref{64}).

{
\section{Conclusions}\label{sec7}

{

\begin{table}[t!] 
\centering
\begin{tabular}{|c|c|r|}
\hline 
$i$ & $r_1$   &  $r_2$    \\ 
\hline\hline
                    1.41  &       0.96  &      0.90 \\
                    1.91  &       0.75  &      0.72 \\
                    2.41  &       0.55  &      0.56 \\
                    2.91  &       0.44  &      0.49 \\
\hline
\end{tabular}
\caption{The ratios $r_1\equiv [Z_1(\beta=0.38) / Z_1(\beta=0)]^2$
and $r_2\equiv [(Z_1/R_D)(\beta=0.38) / (Z_1/R_D)(\beta=0)]^2$ are
shown in columns 2 and 3, respectively, for the values of $i$, shown
in column 1. Notice that these ratios are the same if we use either
$Z_1$ and $R_D$ or $z_1$ and $r_D$.}
\label{tab2}
\end{table}}

We have obtained the generalized impedances $Z_k$ for a range of values of the
two dimensionless parameters $\beta$ and $i$. These allow one to find the current noise
strength from (2.10). We are particularly concerned with the dependence on the
parameter $\beta$ which measures the importance of the capacitance of the junction.
For $\beta=0$ the Langevin equation (1.4) has a simple analytic solution.
An important question is how far the results for $\beta=0$ provide a good
approximation when $\beta$ is of order $0.5$.  
Figures 2 to 8 and table \ref{tab2} exhibit the dependence on $\beta$,
which leads to significant corrections to the voltage noise.

Since the analysis used in \cite{koch} neglected the $\beta$ dependence 
\footnote{
In \cite{koch} it is actually assumed that equation \eqref{29} 
is a good approximation with
observed values of $R_D$ and $\bar{V}$, whereas we have used the theoretical values.
From Fig. 3 of \cite{koch}, it seems that the measured value of $R_D$ in the important range is consistent
with the theoretical value to within about $10\,\%$, but from the data in the published paper
we cannot be more precise about this.}
the conclusions of that paper may not be as decisive as claimed. If similar experiments are done in the
future, our results should be useful.
}

\vfill\eject

\section*{Acknowledgements}
F.T.B. and J.F. would like to thank CNPq, Brazil, for a grant.

\vfill\eject



\end{document}